\documentclass[11pt,twoside]{article}
\usepackage{asp2010}

\resetcounters

\begin{document}

\aspcpryear{2010}
\aspvolauthor{Davenport, et. al}
\aspvoltitle{16th Cambridge Workshop on Cool Stars, Stellar Systems, and the Sun}
\aspvolume{-}

\title{Mining Databases for M Dwarf Variability}
\author{James R.~A. Davenport$^1$, Andrew C. Becker$^1$, Suzanne L. Hawley$^1$, Adam F. Kowalski$^1$, Branimir Sesar$^2$, Roc M. Cutri$^2$
\affil{$^1$Department of Astronomy, University of Washington, Box 351580, Seattle, WA 98195, USA}
\affil{$^2$Infrared Processing and Analysis Center, California Institute of Technology MS 100-22, Pasadena, CA 91125, USA}}

\begin{abstract}
Time-resolved databases with large spatial coverage are quickly becoming a standard tool for all types of astronomical studies. We report preliminary results from our search for %low-mass star variability 
stellar flares in %two time resolved databases, the Lincoln Near-Earth Asteroid Research (LINEAR) project, and 
the 2MASS calibration fields. %The expected frequency of flares from M dwarfs is still an open question, as well as the possibility of flares being seen in the near IR regime. 
A sample of 4343 M dwarfs, spatially matched between the SDSS and the 2MASS calibration fields, each with hundreds to thousands of epochs in near infrared bandpasses, is %presented. 
analyzed using a modified Welch-Stetson index to characterize the variability.  %We search for flares in this combined database, and compare the variability to that seen in recent SDSS Stripe 82 optical studies.
A Monte Carlo model was used to assess the noise of the variability index. 
We find significnat residuals above the noise with power-law slopes of -3.37 and -4.05 for our $JH$ and $HK_s$ distributions respectively.  This is evidence for flares being observed from M dwarfs in infrared photometry.
\end{abstract}

\section{Introduction}
Time domain studies of M dwarfs, using both photometry and spectroscopy, are a critical tool for understanding the nature of magnetically driven flares.% and surface variability.
  The frequency, intensity, and duration of flares are fundamentally tied to the star's ability to release magnetic energy generated by the internal dynamo. Flare events occur stochastically, %with no known forewarning. 
 and therefore large samples are needed to shed light on the physical mechanisms %at work 
 which govern the flare rate, %and 
 its evolution over time, and is variation with stellar %as a function of 
 mass.%, large samples of data are needed to robustly sample the flaring duty cycle of many stars.

%By the mid 1970s, several such studies had been undertaken.
Seminal work %in low-mass star flares was done 
was carried out by \cite{1976ApJS...30...85L} in measuring the frequency of flares from %a handful of 
M dwarfs which had been studied for several years in optical light \citep{1974ApJS...29....1M}. They found that flare occurrence rates, measured as frequency of flares versus energy of flare, could be described using a power-law with a slope between -0.4 and -1. Thus  lower energy flares are seen more frequently.

%modern studies (e.g. w/ the SDSS: adam and eric)
%More 
Recently, use of large area surveys has provided a fresh look at the rate, spatial distribution, and lifetimes of M dwarf flares. 
\cite{2009AJ....138..633K} studied the rate of serendipitous flares found in repeat scans of the SDSS field known as Stripe 82, resulting in flare rates congruent with \cite{1976ApJS...30...85L}. \cite{2010arXiv1009.1158H} examined the repeat exposures from the SDSS DR7 spectral database, %to look for flares. 
finding that flares are more commonly seen in the younger stellar population of the disk, and for later spectral types.
% They recover flares and ... Hilton et al. (2010) 

Stellar flares exhibit a variety of lightcurve morphologies. Many last only seconds to minutes, with a characteristic fast rise and exponential decay, and relatively low amplitudes.  Others have been seen to evolve over many hours, with enormous increases in flux over a wide range in wavelength, and complex lightcurve evolution %structure with several sub-peaks 
\citep[e.g.][]{2010ApJ...714L..98K}.

%Large astronomical databases are growing, with 
The next generation of large photometric surveys will be coming online in the near future. 
Work by Hilton et al. (these proceedings) predicts that M dwarf flares will be observed in nearly every frame of LSST observations, %. These estimates indicate
suggesting that flare stars will be a large source of contamination for other variability studies. 

Current photometric studies to measure rates of M dwarf flaring and magnetic activity have almost exclusively focused on bluer wavelengths. This is understandable as the %typical first-order model of a flare SED is 
flare continuum emission may be modeled with an  8,500K -- 10,000K blackbody \citep{1992ApJS...78..565H, 2010ApJ...714L..98K}, and thus the greatest flux increases will be seen in the shorter wavelength optical bands. However, flares are %magnetic activity is 
known to produce emission over a broad wavelength range % at a wide range of energies 
on the Sun, and it is possible that these events %will also be observed at longer wavelengths. 
may also be observed in the infrared, albeit with reduced amplitude.
%We thus attempted to 
Here we statistically characterize the stochastic increases in flux from M dwarfs at these longer wavelengths and attempt to identify flares.

\section{Sample Selection}
%Our sample consists of spatially matched photometry from three very different surveys. The Lincoln Near-Earth Asteroid Research (LINEAR) Survey was a shallow depth, wide area, visible-light photometric survey which covered a contiguous region in the sky roughly congruent with the SDSS. The LINEAR survey has been complied into a queryable database, and is described in detail in Sesar et. al (in prep).  %This database has been used to search for periodic variables such as eclipsing binaries (Becker et al. 2010)

Our sample consists of spatially matched photometry from two different surveys. 
The 2MASS Calibration Point  Source Working Database (hereafter 2MASScal) consists of photometry from the repeat calibration scans of 40 tiles  spread across the sky, taken as part of the 2MASS survey \citep{2006AJ....131.1163S}. This database is fully described in the 2MASS Explanatory Supplement.\footnote{\url{http://www.ipac.caltech.edu/2mass/releases/allsky/doc/seca4_1.html}} A calibration field was imaged multiple times every night %of the main 2MASS survey 
to ensure repeatable and homogenous photometry for the main survey.
% LINEAR provides roughly hundreds of epochs for its sources, while 
 2MASScal provides between 562 and 3692 scans of each of the 35 main tiles. The five special tiles in the SMC and LMC were each scanned between 108 and 468 times. These databases are presently being mined, and follow-up study being conducted, for periodic variability from objects such as binaries (Davenport et al. in preparation).

Previous work with the 2MASScal data has %investigated variability in low-mass stars using 
used broad $JHK$ color cuts to select low-mass stars \citep{2008ApJS..175..191P}. To reliably determine the spectral type of low-mass stars, we cross matched 16 of the 2MASScal tiles to the Sloan Digital Sky Survey Data Release 7 \citep{2009ApJS..182..543A} 
using a matching radius of 5''. The left panel of Figure 1 shows the location of all the 2MASScal tiles, as well as the  SDSS footprint. The 16 matched fields are shown as purple stars, %We anticipate several more 2MASScal tiles will be included in the forthcoming SDSS data release. This 
yielding 32,044 individual objects with photometry in both SDSS and 2MASScal.

\begin{figure}[!h]
\plotone{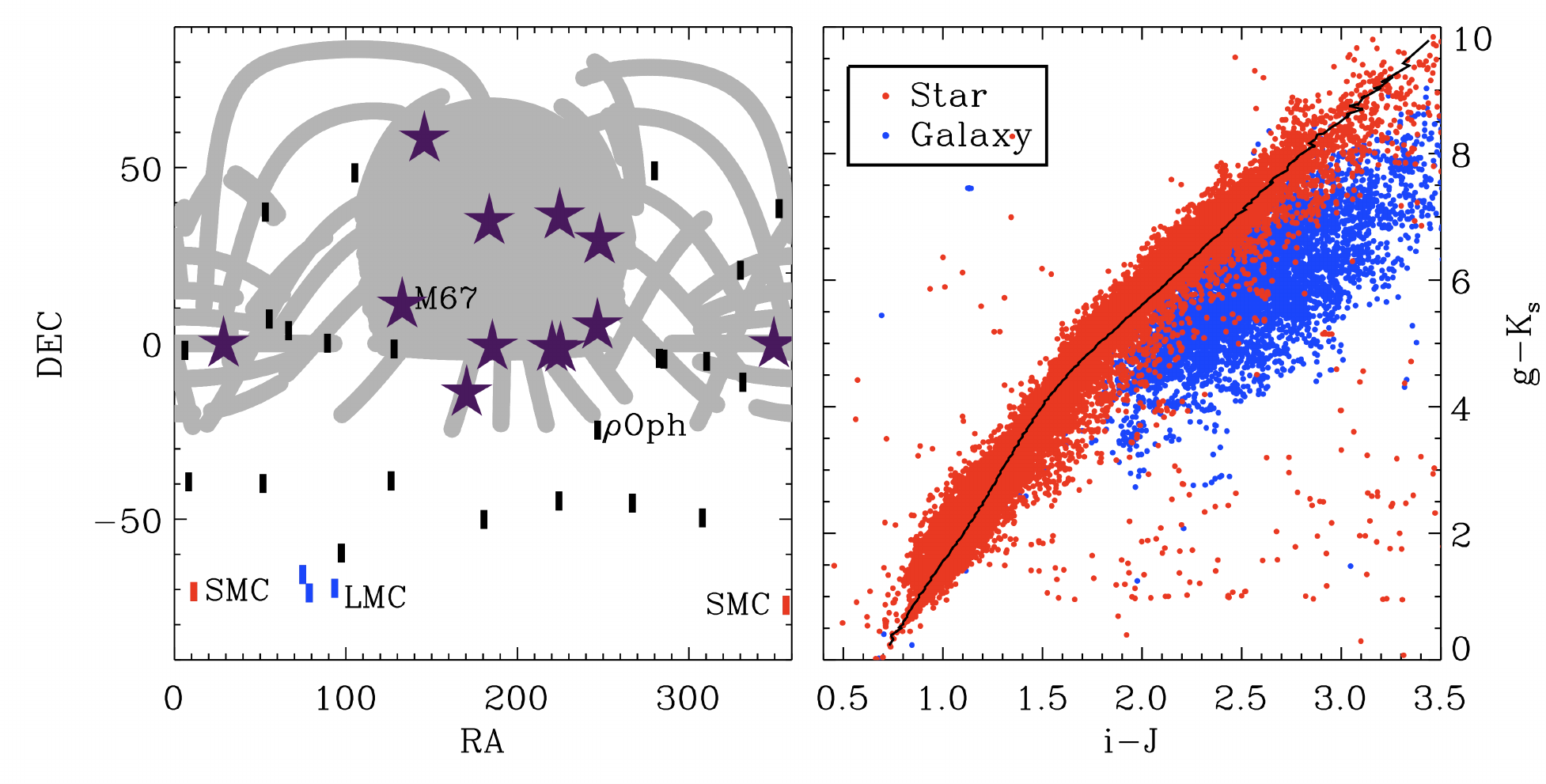}
\caption{(Left) The locations of 2MASScal tiles on the sky. A few notable fields have been labeled. The footprint of the SDSS is shown in grey. Fields with 2MASScal and SDSS photometry are shown as stars. (Right) Color-color diagram of the entire 2MASScal -SDSS matched sample consisting of 24,465 stars, and 7,283 galaxies, as determined by the SDSS \texttt{OBJTYPE} flag. The \cite{2007AJ....134.2398C} fiducial stellar locus is shown in black for reference.}
\end{figure}

The right panel of Figure 1 shows a color--color diagram for the SDSS--2MASScal matched data. For the 2MASScal filters, the mean magnitude for each object over all epochs is used.  Note the large number of galaxies with red colors, a potentially major source of contamination for low-mass star studies. To ensure a clean sample of M dwarfs, we removed any point sources from this matched dataset which the SDSS \texttt{OBJTYPE} flag did not  call a star. Photometric spectral types were then determined with the covariance matrix technique presented in \cite{2009AJ....138..633K} using SDSS $r,i,z$ magnitudes. Because these stars are mostly nearby, the photometry was not corrected for interstellar extinction.
%USE SPTYPE ESTIMATE FROM ADAM
Our final sample, having colors consistent with M0 or later spectral type, contained 4343 stars, with $\sim3.8\times10^6$ 2MASScal photometric measurements.
 %The total number of photometric measurements in 2MASScal for these M dwarfs was 3.77 million.% in 2MASScal, and 0.93 million in LINEAR.

\section{Characterizing Variability}
Finding flares in such a large catalog of observations requires a robust determination of outliers in the stellar lightcurves. In this section we describe the quantities we used to find such epochs, and a model to estimate the reliability of our flare selection.

\subsection{The $\Phi$ Statistic}

Many statistics are available to quantify the variability of stars. %Previous flare studies have made use of 
A popular one is the Welch--Stetson index \citep{1993AJ....105.1813W}. We adopted the modified Welch--Stetson index $\Phi$ used by \cite{2009AJ....138..633K} which was designed to recover flares. %Instead of converting our photometric measurements to fluxes for the unit-less $\Phi$ index, we compute $\Phi$ using magnitudes.  
%We computed $\Phi$ for both $(J,H)$ and $(H,K_s)$. For a given epoch $n$ of a star, 
Since 2MASS observed the $J,H,K_s$ bands simultaneously, at a given epoch $n$, $\Phi_{JH,n}$ is defined as:
\begin{equation}
\Phi_{JH,n} = \left(\frac{m_{J,n}-\bar{m}_J}{\sigma_{J,n}}\right)\left(\frac{m_{H,n}-\bar{m}_H}{\sigma_{H,n}}\right)
\end{equation}
where $m$ is the magnitude at the given epoch, $\bar{m}$ is the mean quiescent magnitude, and $\sigma$ the photometric error at this epoch. We also computed $\Phi_{HK_s,n}$ in the same fashion. The $\Phi$ statistic will 
be positive when both filters are simultaneously brighter or fainter than the mean value, and negative when one is brighter and the other fainter. %Our preliminary analysis presented here is carried out only for  the  $\Phi_{JH}$ and  $\Phi_{LINEAR}$ distributions, and not the $\Phi_{HK_s}$ since the $K_s$ band should show the smallest amplitude response to a flare. 
%For the $\Phi_{JH}$ distribution we cut out
We removed epochs where the star became fainter in at least one band to ensure that we did not recover eclipses or photometric errors such as clouds, but still provided negative $\Phi$ values. Specifically, we eliminated epochs with $m_H > \bar{m}_H$ for both distributions so that they would have the same number of epochs.

%2MASS observes all three $JHK_s$ bands simultaneously \citep{2006AJ....131.1163S}. Because LINEAR makes use of only one clear photometric band, to compute $\Phi_{LINEAR}$ with  we used neighboring epochs which were space very closely in time. A separation of more than an hour was not considered a ``simultaneous'' measurement, and discarded. This reduces the number of points in the  $\Phi_{LINEAR}$ distribution. We arbitrarily chose to cut out  $\Phi_{LINEAR}$ values where the first LINEAR magnitude was below the mean value for the star, to remove epochs with drop-outs or eclipses.

Figure 2 shows the $\Phi_{JH}$ and $\Phi_{HK_s}$ distributions. In both samples, as with the optical results of \cite{2009AJ....138..633K}, the null distribution ($\Phi < 0$) drops off for large values of $\Phi$. 
The $\Phi$ statistic provides a measure of the flux ratio relative to the quiescent value, in terms of the errors for a given epoch. %the number times greater than the photometric error a given epoch's flux difference, relative to the quiescent value, is. 
Flare epoch candidates are selected by defining a cutoff value of $\Phi$ that is characteristic of the noise. Epochs with $\Phi$ greater than this cutoff value, and an increase in brightness as described above, are probable flares.

\begin{figure}[!ht]
\plotone{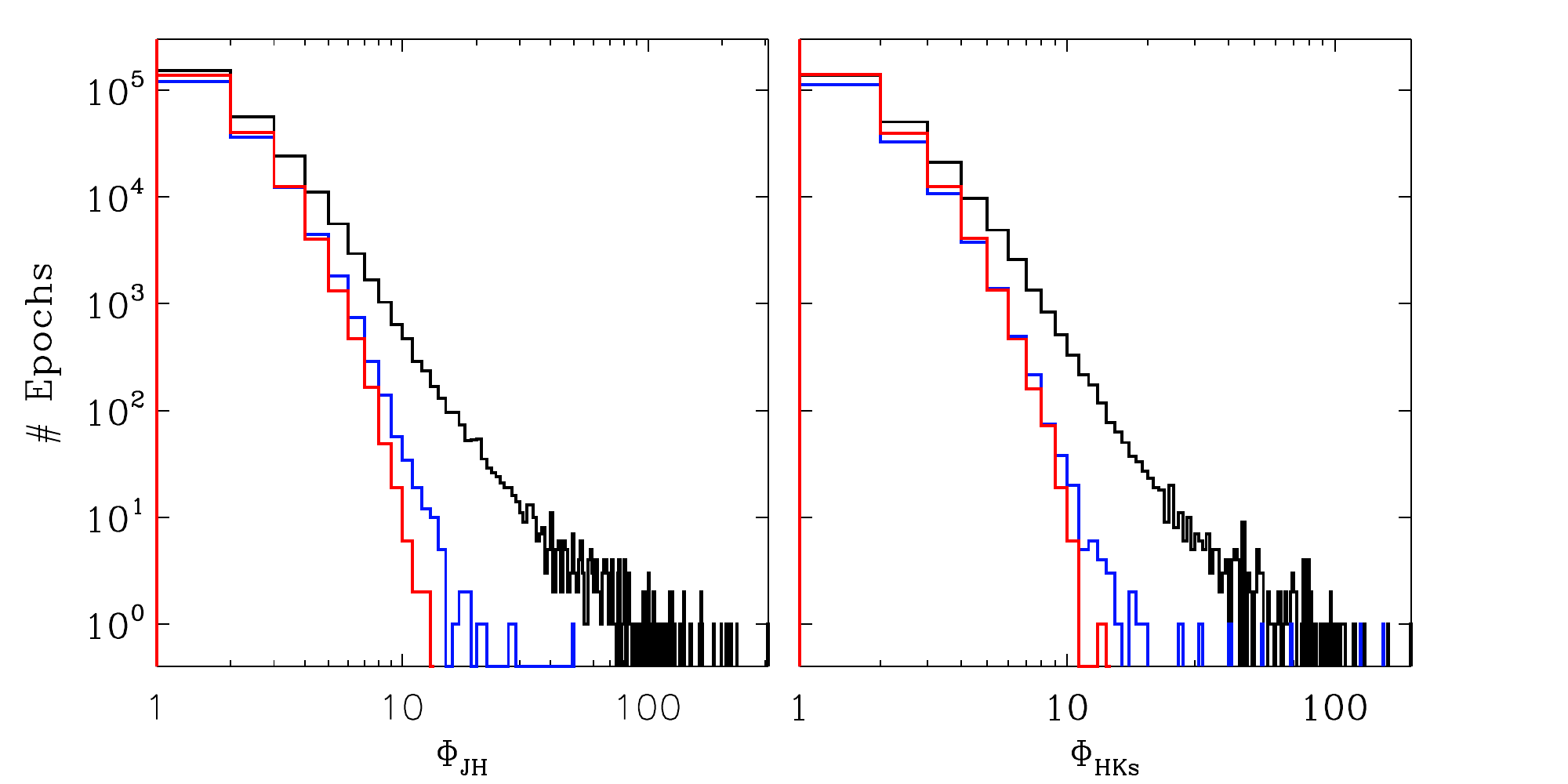}
\caption{Histograms of the flare statistic $\Phi_{JH}$ (Left) and $\Phi_{HK_s}$ (Right). In both panels the black line is the data ($\Phi > 0$) while the blue line is the null sample ($\Phi < 0$). The model null $\Phi$ for each distribution, as described in \S 3.2, is shown in red.}
\end{figure}

%explain the equation, play off of Kowalski's paper. show the $\Phi$ distributions. we need to make a cut. hence we need.....

\subsection{Modeling the Noise}
The null sample of the $\Phi$ distributions, shown in blue in Figure 2, should be representative of the intrinsic statistical noise from the star from low-level stellar variability, or extrinsic sources. To model the noise, we created a Monte Carlo simulation of our dataset. For every star in the sample, random measurements were drawn from a gaussian distribution with a standard deviation equal to the mean photometric error for that object. This was done with the same number of epochs per object as our final catalog contained, thus reproducing the mean properties of the sample.  %The model was run X num of times, producing identical results.

The $\Phi$ statistic was then calculated in the same way as the data using Equation 1 for both $\Phi_{JH}$ and $\Phi_{HK_s}$. Our resulting artificial $\Phi$ distributions were symmetric about $\Phi=0$, as expected for uncorrelated noise between the two bands. We then made the same cuts as in our treatment of the real data to remove $H$--band flux decreases. %epochs with less flux than the mean quiescent flux. 
The model null distributions had approximately half as many epochs as the model quiescent $\Phi>0$ distribution due to these cuts.
This too produced a symmetric $\Phi$ distribution, shown in red in Figure 2, thus allowing us to use the  $\Phi<0$ null sample as a realistic estimate for the quiescent $\Phi>0$ distribution. Any  $\Phi>0$ measurements in excess of the null sample should be good flare candidates.  %The model was run for the  $\Phi_{LINEAR}$ data as well, producing a symmetric null $\Phi$ distribution, as with the 2MASScal model.

%we model the noise to assure ourselves that where it drops off, so too does the quiescent variability which we can't discern. make gaussian model, use intelligent source model

\section{Results}% and Future Work}
We subtracted the real null distributions, shown in Figure 2 in blue, from the data ($\Phi > 0$). The $\Phi$ residuals in Figure 3 show the number of epochs which have $\Phi$ values in excess of the well--characterized random noise. The residual distributions appear to follow a power-law distribution at high $\Phi$ values, as shown in blue by the least-squares fit to the $\Phi>10$ data in each panel.

At $\Phi$ values less than about 10 the residuals turn over, deviating strongly from the single power law slope. This is the regime where the noise dominates, as shown in the model $\Phi$ distributions from \S3.2, and no complete census of very low amplitude deviations is possible. 

\begin{figure}[!ht]
%\plottwo{phi_diff}{phi_diff_linear}
\plotone{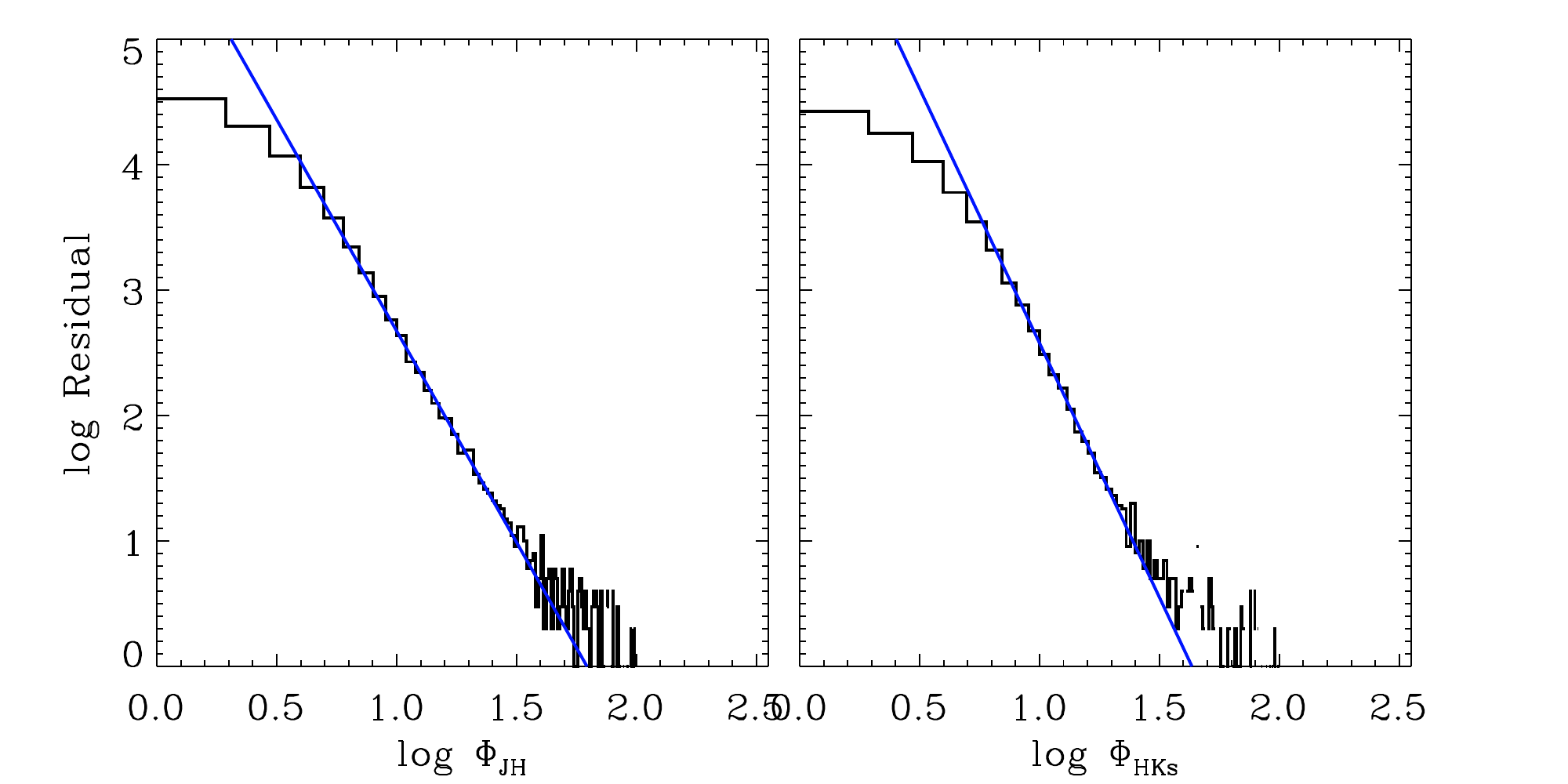}
\caption{The residual $\Phi$ distributions (data - null) for the 2MASScal (Left) and LINEAR (Right)  samples. The blue lines are rough power-law fits to the residual distributions.}
\end{figure}

Our power-law fits have slopes of $-3.37\pm0.12$ and $-4.05 \pm 0.07$ for the  $\Phi_{JH}$ and $\Phi_{HK_s}$ samples, respectively. We caution that these are rough estimates, as these residual distributions contain spectral types M0 and later. It is known that the rate of flares and activity lifetimes vary across the M dwarf spectral sequence \citep{2008AJ....135..785W} and ideally these fits should be done for each spectral type individually. The residual fits presented here are not the same as the power-law fits used in \cite{1976ApJS...30...85L} or \cite{2009AJ....138..633K}, which estimated the energy of the flare and the occurrence frequency. 
However, the trend of the residual distributions seen in Figure 3 suggest that the number of detectable flares decreases with increasing wavelength. %as a function of wavelength suggests that power-law slopes perviously reported in \cite{1976ApJS...30...85L} and others (e.g. see Hilton et al. in these proceedings) steepens as a function of wavelength.
 This is not unexpected, since only the largest amplitude blue flares will have enough flux in the red and infrared bands to be detected.

%====>>>>>>>      COMPARE TO ADAM's STRIPE 82 VALUES!!!

%Flare rate, variability power-laws. give some discussion here, maybe try to compare to some Kowalski \& Hilton work (?)

\section{Future Work and Conclusions}
The large numbers of epochs for most of the sample (typically a couple thousand epochs per object) presents challenges in eliminating source confusion due to varying observing conditions and seeing. Our preliminary investigations of objects with nearest--neighbors closer than 2 arcseconds  show that the 2MASS photometry in our database does not always properly separate objects. In the worst case, some ``objects'' show bimodal distributions in their (RA,Dec) positions. While rare, this can result in objects blending under poor observing conditions, and having abnormally high flux values measured, which in turn can give false positives for flares, and may be the cause of the very low amplitude tail in the $\Phi$ residual plots. 
More stringent tests for the spatial matching and removal or de-blending of such objects are underway and will be presented in future work with this dataset. Our future work will also include matching to the SDSS Stripe 82 time domain photometry, to provide variability analysis for stars in up to eight photometric bands.

We have shown here that statistically significant transient luminosity increases in late type stars, with spectral types M0 and later, are present in NIR survey photometry. The power-law slopes become more  negative at longer infrared wavelengths. This provides some of the first solid evidence for flares appearing in infrared bands, and suggests that flare detection is increasingly rare at longer wavelengths.

\acknowledgements
Funding for the SDSS and SDSS-II has been provided by the Alfred P. Sloan Foundation, the Participating Institutions, the National Science Foundation, the U.S. Department of Energy, the National Aeronautics and Space Administration, the Japanese Monbukagakusho, the Max Planck Society, and the Higher Education Funding Council for England. The SDSS Web Site is http://www.sdss.org/.

The SDSS is managed by the Astrophysical Research Consortium for the Participating Institutions. The Participating Institutions are the American Museum of Natural History, Astrophysical Institute Potsdam, University of Basel, University of Cambridge, Case Western Reserve University, University of Chicago, Drexel University, Fermilab, the Institute for Advanced Study, the Japan Participation Group, Johns Hopkins University, the Joint Institute for Nuclear Astrophysics, the Kavli Institute for Particle Astrophysics and Cosmology, the Korean Scientist Group, the Chinese Academy of Sciences (LAMOST), Los Alamos National Laboratory, the Max-Planck-Institute for Astronomy (MPIA), the Max-Planck-Institute for Astrophysics (MPA), New Mexico State University, Ohio State University, University of Pittsburgh, University of Portsmouth, Princeton University, the United States Naval Observatory, and the University of Washington.

This publication makes use of data products from the Two Micron All Sky Survey, which is a joint project of the University of Massachusetts and the Infrared Processing and Analysis Center/California Institute of Technology, funded by the National Aeronautics and Space Administration and the National Science Foundation.

%future work includes match to spectra to try and understand active vs inactive stars, try to refine the rate as a function of spectral type ,

%%%%%%%%%%%%%%%%%%%%%%%
\bibliography{davenport_j}
%\section{References}
%Kowalski
%Lacy Moffit Evans
%Hilton
%Hawley
%Brani LINEAR
%Becker catalog (submitted)
%Davenport dM Binary (In Prep)

\end{document}